\documentclass{article}
\usepackage{bm}
\usepackage{amsmath,amssymb}
\usepackage[top=20truemm,bottom=20truemm,left=20truemm,right=20truemm]{geometry}
\usepackage{authblk}
\usepackage[backend=biber,style=phys,articletitle=false,biblabel=brackets,chaptertitle=false,pageranges=false,uniquelist=false]{biblatex}
\usepackage{hyperref}
\usepackage{xurl}
\hypersetup{breaklinks=true}
\DeclareSourcemap{
	\maps[datatype=bibtex, overwrite=true]{
		\map{
			\step[fieldset=doi, null]
			\step[fieldsource=journal,
				match={Physical Review D},
				replace={Phys. Rev. D}
			]
			\step[fieldsource=journal,
				match={Philosophical Transactions of the Royal Society A},
				replace={Philos. T. R. Soc. A}
			]
			\step[fieldsource=journal,
				match={The Astrophysical Journal Letters},
				replace={Astrophys. J. Lett.}
			]
			\step[fieldsource=journal,
				match={The Astrophysical Journal},
				replace={Astrophys. J.}
			]
			\step[fieldsource=journal,
				match={Classical and Quantum Gravity},
				replace={Classical Quant. Grav.}
			]
			\step[fieldsource=journal,
				match={Journal of High Energy Physics},
				replace={JHEP}
			]
			\step[fieldsource=journal,
				match={Proceedings of the American Mathematical Society},
				replace={P. Am. Math. Soc.}
			]
			\step[fieldsource=journal,
				match={Journal of Physics A: Mathematical and Theoretical},
				replace={J. Phys. A-Math. Theor.}
			]
			\step[fieldsource=journal,
				match={Journal of Physics A: Mathematical and General},
				replace={J. Phys. A-Math. Gen.}
			]
			\step[fieldsource=journal,
				match={Inventiones mathematicae},
				replace={Invent. Math.}
			]
			\step[fieldsource=journal,
				match={Journal of Geometry and Physics},
				replace={J. Geom. Phys.}
			]
			\step[fieldsource=journal,
				match={Journal of the Faculty of Science, the University of Tokyo. Sect. 1 A, Mathematics},
				replace={J. Fac. Sci. U. Tokyo 1}
			]
			\step[fieldsource=journal,
				match={Compositio Mathematica},
				replace={Compos. Math.}
			]
			\step[fieldsource=journal,
				match={Proceedings of the Japan Academy, Series A, Mathematical Sciences},
				replace={P. Jpn. Acad. A-Math.}
			]
			\step[fieldsource=journal,
				match={Journal of Computational and Applied Mathematics},
				replace={J. Comput. Appl. Math.}
			]
			\step[fieldsource=journal,
				match={Constructive Approximation},
				replace={Constr. Approx.}
			]
			\step[fieldsource=journal,
				match={The European Physical Journal C},
				replace={Eur. Phys. J. C}
			]
			\step[fieldsource=journal,
				match={General Relativity and Gravitation},
				replace={Gen. Relativ. Gravit.}
			]
		}
	}
}
\newcommand{\hgarg}[2]{\genfrac{}{}{0pt}{0}{#1}{#2}}
\makeatletter
	
		\@addtoreset{equation}{section}
\makeatother
\title{Deflection of Light by a Reissner-Nordstr{\"{o}}m Black Hole and Painlev\'e VI equation}
\date{\today}
\author{Tadashi Sasaki\thanks{ta-sasaki@kumagaku.ac.jp}}
\affil{Faculty of Commerce, Kumamoto Gakuen University, Kumamoto 862-8680, Japan}
\begin{document}

\maketitle
\begin{abstract}
%%{{{
	We consider the bending angle of the trajectory of a photon incident from and deflected to infinity around a Reissner-Nordstr{\"{o}}m black hole.
	We treat the bending angle as a function of the squared reciprocal of the impact parameter and the squared electric charge of the background normalized by the mass of the black hole.
	It is shown that the bending angle satisfies a system of two inhomogeneous linear partial differential equations with polynomial coefficients.
	This system can be understood as an isomonodromic deformation of the inhomogeneous Picard-Fuchs equation satisfied by the bending angle in the Schwarzschild spacetime,
	where the deformation parameter is identified as the background electric charge. 
	Furthermore, the integrability condition for these equations is found to be a specific type of the Painlev\'e VI equation that allows an algebraic solution.
	We solve the differential equations both at the weak and strong deflection limits.
	In the weak deflection limit, the bending angle is expressed as a power series expansion in terms of the squared reciprocal of the impact parameter and
	we obtain the explicit full-order expression for the coefficients.
	In the strong deflection limit, we obtain the asymptotic form of the bending angle that consists of the divergent logarithmic term and the finite $O(1)$ term
	supplemented by linear recurrence relations which enable us to straightforwardly derive higher order coefficients.
	In deriving these results, the isomonodromic property of the differential equations plays an important role.
	Lastly, we briefly discuss the applicability of our method to other types of spacetimes such as a spinning black hole.
%%}}}
\end{abstract}

\tableofcontents

\section{Introduction}
%%{{{
Light deflection by a massive object has been considered one of the remarkable successes of general relativity since the confirmation during a total solar eclipse in 1919\cite{Dyson_1920}.
After the first direct observation of gravitational wave signals from a binary black hole merger\cite{LIGO2016},
the expectation for the confirmation of general relativity or alternative gravitational theories in the strong field regime has been increasing. 
In this context, the importance of observations by electromagnetic waves is no less than before, 
as, for example,  the Event Horizon Telescope collaboration revealed the existence of black hole shadows\cite{EHT_2019,EHT_2022}.

Propagation of electromagnetic waves on a given background spacetime can be described by null geodesic equations in the geometric optics approximation.
Geodesic equations reflect the metric of the background spacetime therefore observing trajectories of photons that pass close to a massive compact object,
such as a black hole, helps us understand spacetime geometry in a strong gravitational field.
One of the fundamental quantities associated with a photon's trajectory is its bending angle $\alpha$. 
Note that the notion of the bending angle is valid for trajectories in spherically symmetric spacetimes or on the equatorial plane in axisymmetric spacetimes. 
Otherwise one has to integrate null geodesic equations for all coordinates\cite{VazquezEsteban2003,Sereno_2006,Bozza_2006,Kraniotis_2011,GrallaLupsasca2020a,GrallaLupsasca2020b,Hou_2022}.
In this article, we assume both the source and the observer lie sufficiently far from the lensing object
although we give a remark on applicability to the cases without this restriction.

The bending angle has been studied both analytically and numerically on various spacetime metrics.
When the photon's trajectory is sufficiently distant from the lensing object, 
the spacetime metric around the trajectory can be approximated by the Minkowski metric with linear perturbations.
In this case, the bending angle is proportional to the perturbations and is considered small.
The leading behavior of the bending angle $\alpha$ in this limit is given by
\begin{equation}
	\alpha\sim\frac{4M}{b}, \label{EinsteinFormula}
\end{equation}
where $M$ and $b$ represent the mass of the lensing body and the impact parameter of the photon's trajectory respectively, and $M/b\ll1$ is assumed.
Note that we employ the geometrized units where the speed of light $c$ in vacuum and the gravitational constant $G$ are set equal to unity in this article.
When the photon's trajectory becomes closer to the lensing object, this formula is modified by higher order terms in $M/b$,
which give a power series expression for the bending angle.
However, the convergence radius of this power series is usually finite and the bending angle diverges at some critical value of $b$,
which implies that photons can go around once or more the lensing body before escaping to infinity.
This regime is called the strong deflection limit, where the bending angle is typically approximated by the following form:
\begin{equation}
	\alpha\sim-\bar{a}\log\left(\frac{b}{b_c}-1\right)+\bar{b},\label{BozzaFormula}
\end{equation}
where $\bar{a}$ and $\bar{b}$ are constants depending only on the parameters of the background spacetime and $b_c$ is the critical value of the impact parameter\cite{Bozza_2002}.

In this article, we consider the problem of calculating the bending angle in the Reissner-Nordstr{\"{o}}m spacetime.
Gravitational lensing in the Reissner-Nordstr{\"{o}}m spacetime was first studied in \cite{Eiroa_2002}. 
They expressed the bending angle in terms of the incomplete elliptic integral of the first kind,
from which approximation for the bending angle was derived in terms of the closest approach distance instead of the impact parameter.
For the weak deflection limit, the expansion coefficients were calculated up to the 2nd order while for the strong deflection limit,
only some numerical values of the leading coefficients were shown.
Analytical calculations for the weak deflection limit proceeded in \cite{Keeton_2005},
where the weak deflection expansion was calculated through the PPN formalism and the Reissner-Nordstr{\"{o}}m metric was treated as an example.
They derived explicit coefficients as a function of the background charge up to $O((M/b)^3)$-term.
On the other hand, the strong deflection limit for a generic spherically symmetric spacetime was studied in \cite{Bozza_2002}
and the divergence in this limit was shown to be logarithmic in general.
The author applied the result to the Reissner-Nordstr{\"{o}}m spacetime as an example and obtained analytic expressions that are valid up to the second order in the background electric charge.
Analytical calculation of the bending angle for the Reissner-Nordstr{\"{o}}m spacetime in the strong deflection limit was improved in \cite{TsukamotoGong_2017},
where the analytic expression up to the nonvanishing term was explicitly obtained.
The method was generalized to an arbitrary asymptotically flat, static, spherically symmetric spacetime in\cite{Tsukamoto_2017}.

One of the objectives of this article is to proceed with analytical calculation to derive a full-order expression for the bending angle.
The method we employ in this article is an extension of that in \cite{Sasaki_2021}, 
where the bending angle in the Schwarzschild and the extremal Reissner-Nordstr{\"{o}}m spacetimes were analytically calculated
by solving the inhomogeneous Picard-Fuchs differential equations.
The Picard-Fuchs equations are differential equations satisfied by the period integrals on algebraic manifolds.
In \cite{Sasaki_2021}, the bending angle is regarded as a period integral on elliptic curves defined by a third-order polynomial,
one of whose roots defines the closest approach distance of the photon's trajectory.
As a function of the reciprocal of the impact parameter normalized by the mass of the lensing black hole, 
the Picard-Fuchs equation was shown to be a Gauss hypergeometric equation with an inhomogeneous term in each case,
which enables us to obtain full-order analytic expressions for both the weak and the strong deflection limits.

As we see in Section \ref{PFeq}, a major difference from the cases studied in \cite{Sasaki_2021} is that
the bending angle obeys a system of two linear partial differential equations while 
a single ordinary differential equation appeared in \cite{Sasaki_2021}.
An additional differential equation describes the dependence of the bending angle with respect to the background electric charge.
Solving this system, we derive analytic expressions for the bending angle as a function of the impact parameter and the background parameters.

Another objective is to point out that the system of differential equations can be regarded as an isomonodromic deformation
of the ordinary differential equation for the cases of the Schwarzschild and the extremal Reissner-Nordstr{\"{o}}m spacetimes
and a specific type of the Painlev\'e VI equations appears as the integrability condition for the system.
We note that recently the notion of isomonodromy has been used to study scattering problems on various spacetimes representing black holes\cite{NovaesCunha2014,Cunha2015,NovaesCunha2015,NovaesCunha2016,AmadoCunhaPallante2020,CunhaCavalcante2021,CavalcanteCunha2021,AmadoCunhaPallante2022}, where various types of the Painlev\'e equations and associated $\tau$-functions have been used to analyze the quasinormal modes of linear perturbations.

This article is organized as follows: in Section \ref{geodesic}, we derive an integral representation of the deflection angle from the first integrals of null geodesics.
In Section \ref{PFeq}, we analyze the bending angle as a function of the background parameters and the impact parameter of the trajectory
through the inhomogeneous Picard-Fuchs equations, which are derived in subsection \ref{derivePFeq}.
Subsections \ref{homsolsection} and \ref{inhomsolsection} are devoted to examining homogeneous and inhomogeneous solutions to these equations respectively.
In particular, we obtain full-order analytic expressions for the power series expansions around the weak deflection limit
while for the strong deflection limit linear recurrence relations of the expansion coefficients are given.
The results in these subsections are combined to obtain analytic expressions for the bending angle around the weak and strong deflection limits in subsection \ref{phisolsection}.
In subsection \ref{PVI-PF}, we show the relationship between the Picard-Fuchs equations considered above and the Painlev\'e VI equation.
We summarize this article and give some discussion in Section \ref{summary}.

%%}}}
\section{Integral Representation of Deflection Angle\label{geodesic}}
%%{{{1
The Reissner-Nordstr{\"{o}}m metric is a solution to the vacuum Einstein-Maxwell equation, 
which represents an electrically charged spherically symmetric, and static spacetime.
It can be written in the following form with a suitable choice of a coordinate system:
\begin{equation}
	ds^2=-\left(1-\frac{2M}{r}+\frac{Q^2}{r^2}\right)dt^2+\left(1-\frac{2M}{r}+\frac{Q^2}{r^2}\right)^{-1}dr^2
		+r^2d\theta^2+r^2\sin^2\theta d\phi^2,
\end{equation}
where $M$ and $Q$ correspond to the mass and the electric charge of the spacetime respectively.
We assume $M>0$ and $|Q|\leq M$, otherwise the spacetime has a naked singularity.
When $Q=0$ the metric represents a Schwarzschild black hole while it is called extremal when $|Q|=M$.
The spacetime has two horizons at $r=r_\pm^{(H)}=M\pm\sqrt{M^2-Q^2}$ and the outer horizon corresponds to the event horizon.

Trajectories of photons are described by the null geodesic equations with the above metric.
Since the spacetime is spherically symmetric and static, any null geodesic is confined in a plane defined by the initial tangent vector of the trajectory and the radial direction.
We can set the coordinates so that the plane of the trajectory corresponds to the equatorial plane $(\theta=\pi/2)$ without loss of generality.
Then we have the following expressions for the first integrals of motion for the trajectory:
\begin{align}
	\epsilon&=\left(1-\frac{2M}{r}+\frac{Q^2}{r^2}\right)\frac{dt}{d\lambda}, \\
	L&=r^2\frac{d\phi}{d\lambda},
\end{align}
where $\epsilon\ (>0)$ and $L$ are constants corresponding to the energy and the angular momentum of the photon, 
and $\lambda$ is an affine parameter of the trajectory.
Null geodesics are described by combining these equations and the null condition $(ds/d\lambda)^2=0$.
By eliminating $\lambda$, we obtain 
\begin{equation}
	\left(\frac{d\phi}{dr}\right)^2=\frac{L^2/r^4}{\epsilon^2-(L^2/r^2)(1-2M/r+Q^2/r^2)}.
\end{equation}
Defining the dimensionless parameters as $u=2M/r$, $y=Q^2/4M^2$, and $z=4M^2\epsilon^2/L^2$, this equation reduces to 
\begin{equation}
	\left(\frac{d\phi}{du}\right)^2=\frac{1}{-yu^4+u^3-u^2+z}.\label{dphi/du}
\end{equation}
Note that the parameter $z$ is related to the impact parameter $b$ of the trajectory as $z=4M^2/b^2$.

We now clarify the physically relevant region of the parameters $y$ and $z$.
The requirement $0\leq|Q|\leq M$ mentioned above is translated into $0\leq y \leq 1/4$.
In this article, we consider trajectories of photons coming from and going back to infinity $(u=0)$.
The position of the turning point of such a trajectory is given by the smallest positive root of the quartic polynomial $-yu^4+u^3-u^2+z$.
For a fixed value of $y$, the quartic polynomial $f(u):=yu^4-u^3+u^2$ has local minima at $u=0$ and $u=u_+$, and a local maximum at $u=u_-$,
where $u_\pm=(3\pm\sqrt{9-32y})/8y$.
Note that $u_-$ corresponds to the radius of the photon sphere.
Zeros of $f(u)$ are $u=0$ and $u=u_\pm^{(H)}=(1\pm\sqrt{1-4y})/2y$, the latter of which correspond to the inner and outer horizons.
Under the restriction $0\leq y\leq 1/4$, we have $u_-<u_-^{(H)}\leq u_+\leq u_+^{(H)}$.
Therefore, the physical parameter region for $z$ is given by $0\leq z<f(u_-)$, otherwise the smallest positive root of $f(u)=z$ lies inside the event horizon.
We denote the upper limit $f(u_-)$ of $z$ by $z_+$ for a fixed $y$, which is explicitly given by
\begin{equation}
	z_+(y)=\frac{-128y^2+144y-27+(9-32y)^{3/2}}{512y^3}. \label{zplus}
\end{equation}

We denote the poles of the right-hand side of eq.(\ref{dphi/du}) by $u_i\ (i=1,2,3,4)$.
In the parameter region $0\leq y\leq 1/4$ and $0\leq z<z_+$, only one of them are non-positive 
so we can take $u_1\leq 0\leq u_2\leq u_3\leq u_4$ without loss of generality.
Since the turning point of the trajectory corresponds to $u=u_2$,
the integral representation of the deflection angle $\Delta\phi$ is given by integrating eq.(\ref{dphi/du}) from $u=0$ to $u=u_2$ 
\footnote{Precisely speaking, "deflection angle" usually refers to $\alpha=2\Delta\phi-\pi$ in the literature, 
and $\Delta\phi$ corresponds to the change of the azimuthal angle between infinity and the turning point.
But we call $\Delta\phi$ deflection angle for brevity.},
\begin{equation}
	\Delta\phi=\int_0^{u_2}\frac{du}{\sqrt{-yu^4+u^3-u^2+z}}=\int_0^{u_2}\frac{du}{\sqrt{y(u-u_1)(u_2-u)(u_3-u)(u_4-u)}}. \label{intexpphi}
\end{equation}

As one can see from the last expression, $\Delta\phi$ is an incomplete elliptic integral.
Therefore we can in principle write $\Delta\phi$ in terms of the standard elliptic integrals such as Legendre's ones\cite{Eiroa_2002}.
However, such expressions are not so illuminating to understand $\Delta\phi$ as a function of $y$ and $z$
because the modulus of the elliptic integral is a complicated function of these parameters.

What we attempt to do in this article is to understand the dependence of $\Delta\phi$ on $y$ and $z$.
In particular, there are two physically important asymptotic regions of parameter $z$, 
i.e. the weak deflection limit $z\to0$ and the strong deflection limit $z\to z_+(y)$ with $y$ kept fixed.
The reason why there exists an upper limit of $z$ stems from the existence of the circular orbits of photons in this spacetime.
As $z$ increases from $0$, the turning point of the trajectory gets closer to the circular orbits and the winding number also increases.
In the strong deflection limit $z\to z_+(y)$, the photon is trapped in that orbit and never goes back to infinity.
Mathematically speaking, $\Delta\phi$ diverges as $z$ approaches $z_+(y)$
since the roots $u_2$ and $u_3$ merge in this limit to give a simple pole in the integrand of eq.(\ref{intexpphi}).
In particular, we see the divergence of the deflection angle is logarithmic.
For later convenience, we introduce $s=\sqrt{1-32y/9}$, where the physical interval $0\leq y\leq1/4$ translates into $1/3\leq s\leq1$.
Using this parameter, we have $z_+(s)=8(3s+1)/27(s+1)^3$.

%%}}}
\section{Inhomogeneous Picard-Fuchs Equation for the Deflection Angle\label{PFeq}}
%%{{{

In this section, we derive the inhomogeneous Picard-Fuchs differential equations (\ref{zeqphi}) and (\ref{yeqphi}) satisfied by $\Delta\phi$ as a function of $y$ and $z$,
and solve them to understand the behavior of $\Delta\phi$ both in the weak and strong deflection regimes.
In subsection \ref{derivePFeq}, we describe the derivation of the differential equations for $\Delta\phi$ 
and show that the differential operator for eq.(\ref{zeqphi}) is a Fuchsian one, i.e. all the singularities are regular ones,
which include the weak ($z=0$) and strong ($z=z_+$) deflection limits.
In subsections \ref{homsolsection}, \ref{inhomsolsection}, and \ref{phisolsection}, we solve the differential equations for $\Delta\phi$.
Lastly in subsection \ref{PVI-PF}, we discuss the relationship of the inhomogeneous Picard-Fuchs equations for $\Delta\phi$ and isomonodromic deformations of Fuchsian differential equations.
In particular, we show that the integrability condition for our system corresponds to the Painlev\'e VI equation with a specific set of parameters,
where the Painlev\'e equation allows an algebraic solution.

	\subsection{Derivation of the Inhomogeneous Picard-Fuchs Equation\label{derivePFeq}}
	%%{{{
	The method to derive the differential equations for $\Delta\phi$ is an extension of that used in \cite{Sasaki_2021}
	for the Schwarzschild and the extremal Reissner-Nordstr{\"{o}}m cases.
	We look for an identity of the following form for some positive integer $N$:
	\begin{equation}
		\left(c_N\partial^N_z+\cdots+c_1\partial_z+c_0\right)K=\partial_u(RK^{2N-1}), \label{PFidgen}
	\end{equation}
	where $K=K(y,z,u)=(-yu^4+u^3-u^2+z)^{-1/2}$ is the integrand of $\Delta\phi$.
	$c_i$ and $R$ are undetermined functions of $(y,z)$ and $(y,z,u)$ respectively.
	Note that $R$ is assumed to be a polynomial in $u$.

	If we integrate the one form $Kdu$ along a path going from one of the roots of $-yu^4+u^3-u^2+z$ to another one and coming back to the initial one
	on the complex $u$-plane,
	we obtain one of the periods of the torus defined by this quartic polynomial.
	Integrating eq.(\ref{PFidgen}) on the same path, the right-hand side will vanish since the path is closed.
	In this way, we can see that the periods $\omega$ of the torus obey a homogeneous differential equation, namely the Picard-Fuchs (PF) equation.

	There are two linearly independent periods, say $\omega_1$ and $\omega_2$, on a torus.
	Accordingly, the differential equation should have two independent solutions.
	Thus, the PF equation on a torus should be a 2nd order one, i.e. $N=2$.
	Assuming $N=2$, we can see that the order of the polynomial $R$ must be $6$ by counting the order of $u$ in eq.(\ref{PFidgen}),
	so we set $R=R_0+R_1u+\cdots +R_6u^6$.
	Inserting this into the identity (\ref{PFidgen}) with $N=2$, we obtain a system of linear equations for undetermined coefficients $c_i$ and $R_j$,
	\begin{equation}
	\begin{pmatrix}
		4z^2 & -2z & 3 & 0 & -4z & 0 & 0 & 0 & 0 & 0 \\
		0 & 0 & 0 & 3 & 0 & 2z & 0 & 0 & 0 & 0 \\
		4z & -1 & 0 & -9 & 4 & 0 & 6z & 0 & 0 & 0 \\
		-4z & 1 & 0 & 12y & -7 & 2 & 0 & 8z & 0 & 0 \\
		4yz-2 & -y & 0 & 0 & 10y & -5 & 0 & 0 & 10z & 0 \\
		4 & 0 & 0 & 0 & 0 & 8y & -3 & -2 & 0 & 12z \\
		4y+2 & 0 & 0 & 0 & 0 & 0 & -6y & 1 & 4 & 0 \\
		4y & 0 & 0 & 0 & 0 & 0 & 0 & 4y & 1 & -6 \\
		-2y^2 & 0 & 0 & 0 & 0 & 0 & 0 & 0 & 2y & 3 
	\end{pmatrix}
	\begin{pmatrix}
		c_0 \\
		c_1 \\
		c_2 \\
		R_0 \\
		R_1 \\
		R_2 \\
		R_3 \\
		R_4 \\
		R_5 \\
		R_6
	\end{pmatrix}=0.
	\end{equation}
	We have 9 equations for 10 coefficients, so there remains one arbitrary coefficient, say $c_0$, which determines the overall normalization.
	Fixing this ambiguity by setting
	\begin{equation}
		c_0=3\left[128y^4(8y-3)z^2+2y(4y-1)(64y^2+12y-3)z+(4y-5)(4y-1)^2\right],
	\end{equation}
	the other coefficients are found to be
	\begin{align}
		c_1&=8\bigl[512y^4(8y-3)z^3+y(2560y^3-1920y^2+648y-81)z^2 \notag\\
		&\hspace{4em}
			+(4y-1)(128y^2-144y+27)z+2(4y-1)^2\bigr], \\
		c_2&=4z\bigl[256y^3z^2+(128y^2-144y+27)z+4(4y-1)\bigr]\bigl[2y(8y-3)z+4y-1\bigr], \label{c2}\\
		R_0&=4z\bigl[8y^3(8y-3)z^2+4y(4y-1)(13y-3)z-(4y-1)^2\bigr], \\
		R_1&=-4\bigl[32y^4(8y-3)z^3-4y(32y^3-150y^2+72y-9)z^2 \notag\\
		&\hspace{4em}
			-(4y-1)(28y^2-3)z-(4y-1)^2\bigr], \\
		R_2&=-6\bigl[8y^3(8y-3)z^2+4y(4y-1)(13y-3)z-(4y-1)^2\bigr], \\
		R_3&=8\bigl[2y^3(8y-3)(32y+3)z^2+y(4y-1)(64y^2+27y-6)z+(4y-1)(8y^2-21y+5)\bigr], \\
		R_4&=-6\bigl[80y^4(8y-3)z^2+2y(4y-1)(52y^2+6y-3)z-(4y-1)(16y^2+14y-5)\bigr], \\
		R_5&=12y\bigl[32y^4(8y-3)z^2+2y(4y-1)(16y^2+9y-3)z+(4y-1)(4y^2-19y+5)\bigr], \\
		R_6&=-2y^2(4y-1)\bigl[6y(8y-3)z-52y+15\bigr].
	\end{align}
	The PF equation for a period $\omega$ of the torus is then given by
	\begin{equation}
		\left(c_2\partial^2_z+c_1\partial_z+c_0\right)\omega=0. \label{homPFz}
	\end{equation}

	On the other hand, the deflection angle $\Delta\phi$ is given by the integration from $u=0$ to $u=u_2$, 
	the former of which is not any of the roots of $-yu^4+u^3-u^2+z$ except for the case of $z=0$.
	Thus, the differential equation for $\Delta\phi$ contains an inhomogeneous term coming from the value of $RK^{2N-1}$ at the boundary $u=0$.
	Integrating eq.(\ref{PFidgen}) from $u=0$ to $u=u_2$, the following differential equation for $\Delta\phi$ is obtained:
	\begin{equation}
		\left(\partial^2_z+p_1\partial_z+p_0\right)\Delta\phi=-\frac{R_0}{c_2z^{3/2}}, \label{zeqphi}
	\end{equation}
	where $p_1=c_1/c_2$ and $p_0=c_0/c_2$.
	We here normalized the coefficient of $\partial^2_z$ to unity so that the singular points of the differential equation can be seen from
	the poles of $p_1$ and $p_0$.

	Note that in deriving this we have to deal with the additional terms coming from the differentiation of the boundary of the integration, i.e. $u_2$.
	At the same time, we notice that each of such terms is divergent at $u=u_2$.
	We regularize the divergence by changing the upper limit of the integral as $u_2-\varepsilon$ with a small positive parameter $\varepsilon$.
	For example, the first derivative $\partial_z\Delta\phi$ is replaced by
	\begin{equation}
		\partial_z\int_0^{u_2-\varepsilon}K(y,z,u)du=\int_0^{u_2-\varepsilon}\partial_zK(y,z,u)du+\partial_zu_2\cdot K(y,z,u_2-\varepsilon),
	\end{equation}
	and after combining all the terms we take the limit $\varepsilon\to+0$.
	In fact, one can show that the divergences of these terms cancel with the divergence coming from the inhomogeneous term $RK^3$ at $u=u_2-\varepsilon$.

	From the expression (\ref{c2}) of $c_2$, we observe that the singularities on the finite complex plane for eq.(\ref{zeqphi}) are $z=0$, $z=\Lambda:=(1-4y)/2y(8y-3)$, and
	\begin{equation}
		z_\pm=\frac{-128y^2+144y-27\pm(9-32y)^{3/2}}{512y^3}=\frac{8(3s\pm1)}{27(s\pm1)^3},
	\end{equation}
	one of which is indeed the critical point for the strong deflection limit, i.e. eq.(\ref{zplus}).
	Note that when $0<y<1/4$, $z_-<\Lambda<0<z_+$ holds. Therefore $z_-$ and $\Lambda$ are outside the physically relevant region.
	Since all the roots of $c_2$ are nondegenerate, these are regular singularities of (\ref{homPFz}).
	In addition, $z=\infty$ is also a regular singularity, which can be confirmed by changing the independent variable as $z\to w=1/z$.
	Thus, eq.(\ref{zeqphi}) is a Fuchsian differential equation with 5 regular singularities.

	The characteristic exponents at these singularities can be seen from the residues of $p_1$ since there is no 2nd-order pole in $p_0$ except for $z=\infty$. 
	Explicitly, $p_1$ is given by
	\begin{equation}
		p_1=\frac{1}{z}+\frac{1}{z-z_+}+\frac{1}{z-z_-}-\frac{1}{z-\Lambda},
	\end{equation}
	which shows that the characteristic exponents at $z=0$ and $z=z_\pm$ are 0 (degenerate), while those are $0$ and $2$ at $z=\Lambda$.
	In general, if the difference of two characteristic exponents at a singularity of some linear ordinary differential equation is an integer,
	one of the local solutions usually has logarithmic terms and is not holomorphic at that point.
	This is indeed the case for $z=0$ and $z=z_\pm$, while any local solutions around $z=\Lambda$ are found to be holomorphic.
	Such a singularity is called an apparent singularity.
	Similarly, the characteristic exponents at $z=\infty$ are found to be $1/4$ and $3/4$.
	We summarize the result so far in the following Riemann scheme:
	\begin{equation}
		\left\{
			\begin{matrix}
				z=0 & z=z_+ & z=z_-& z=\Lambda & z=\infty \\
				0 & 0 & 0 & 0 & 1/4 \\
				0 & 0 & 0 & 2 & 3/4
			\end{matrix}
		\right\}.
	\end{equation}

	Although the $z$-dependence of $\Delta\phi$ is totally determined by (\ref{zeqphi}), the $y$-dependence is not so restricted by this equation.
	In fact, $\Delta\phi$ satisfies another differential equation that contains the $y$-derivative of it as we now show.
	As in the case of deriving (\ref{zeqphi}), we consider an identity for $K$ of the following form:
	\begin{equation}
		\left(b_z\partial_z+b_y\partial_y+b_0\right)K=\partial_u\left(\tilde{R}K\right),
	\end{equation}
	where $b_z$, $b_y$, and $b_0$ are functions of $(y,z)$ and $\tilde{R}$ is a polynomial of $u$ with $(y,z)$-dependent coefficients.
	These are found to be, with a suitable normalization, 
	\begin{align}
		b_z&=2z\left[(8y-9)z+2\right], \\
		b_y&=-2y(8y-3)z-4y+1, \\
		b_0&=-3z, \\
		\tilde{R}&=u^2-2\left[(4y-3)z+1\right]u-2z.
	\end{align}
	Then, the following partial differential equation for $\Delta\phi$ is derived:
	\begin{equation}
		\left(\partial_y+q_1\partial_z+q_0\right)\Delta\phi=\frac{2\sqrt{z}}{b_y}, \label{yeqphi}
	\end{equation}
	where $q_1=b_z/b_y$ and $q_0=b_0/b_y$.
	Finally, we have obtained a system of linear partial differential equations for $\Delta\phi$, namely eq.(\ref{zeqphi}) and (\ref{yeqphi}).

	Before going to the next subsection, we examine the Schwarzschild $(y\to0)$ and extremal $(y\to1/4)$ limits of (\ref{zeqphi}) keeping $z$ finite.
	In each case, one can see that eq.(\ref{zeqphi}) reduces to the Gauss hypergeometric differential equation with an inhomogeneous term,
	\begin{align}
		\left[4z(4-27z)\partial^2_z+8(2-27z)\partial_z-15\right]\left.\Delta\phi\right|_{y=0}&=\frac{4}{\sqrt{z}}, \label{Schwarzschildeq}\\
		\left[4z(1-4z)\partial^2_z+4(1-8z)\partial_z-3\right]\left.\Delta\phi\right|_{y=1/4}&=\frac{1}{\sqrt{z}} \label{eRNeq}.
	\end{align}
	These cases are studied in \cite{Sasaki_2021} and they showed that $\Delta\phi_{y=0,1/4}$ can be written in terms of hypergeometric functions ${}_2F_1$ and ${}_3F_2$.

	%%}}}
	\subsection{Homogeneous Solutions\label{homsolsection}}
	%%{{{
	Before considering the differential equations for $\Delta\phi$ itself, we investigate the homogeneous ones, 
	\begin{align}
		\left(\partial^2_z+p_1\partial_z+p_0\right)\omega&=0, \label{zeqomega}\\
		\left(\partial_y+q_1\partial_z+q_0\right)\omega&=0. \label{yeqomega}
	\end{align}
	As is well known, the solution space of the PF equation consists of all linear combinations of the periods of the torus associated with the polynomial $-yu^4+u^3-u^2+z$.
	The integral expressions for the periods can be constructed by integrating the 1-form $Kdu$ between any two of the roots of this polynomial,
	while at most only two of them can be linearly independent.
	As a basis of the solution space, we take 
	\begin{equation}
		\omega_1=\int_{u_1}^{u_2}\frac{du}{\sqrt{-yu^4+u^3-u^2+z}},\ \ \omega_2=i\int_{u_2}^{u_3}\frac{du}{\sqrt{yu^4-u^3+u^2-z}}.\label{periods}
	\end{equation}
	Note that $\omega_1$ and $\omega_2$ are real and purely imaginary valued respectively in the physically relevant region $0\leq y\leq1/4$ and $0\leq z< z_+$.
	
	Let us consider local solutions of (\ref{zeqomega}) around $z=0$.
	Because the characteristic exponent here is $0$, we can set a basis of local solutions $\omega_r^{(0)}$ and $\omega_l^{(0)}$ as
	\begin{align}
		\omega_r^{(0)}&=\sum_{n=0}^\infty a_n^{(0)}z^n,\label{omega_r0}\\
		\omega_l^{(0)}&=-\omega_r^{(0)}\log z+\sum_{n=1}^\infty b_n^{(0)}z^n,
	\end{align}
	where $a_n^{(0)}$ and $b_n^{(0)}$ are possibly $y$-dependent coefficients and normalized by $a_0^{(0)}=1$.
	From eq.(\ref{zeqomega}), both $a_n^{(0)}$ and $b_n^{(0)}$ are uniquely determined, of which we show the first few coefficients below, 
	\begin{align}
		a_1^{(0)}&=-\frac{3}{16}(4y-5),\ \ a_2^{(0)}=\frac{105}{1024}(16y^2-72y+33),\\
		b_1^{(0)}&=\frac{1}{4(4y-1)}-\frac{37}{8}+\frac{5}{2}y,\ \ b_2^{(0)}=\frac{1}{32(4y-1)^2}+\frac{7}{16(4y-1)}-\frac{18997}{1024}+\frac{4341}{128}y-\frac{389}{64}y^2.
	\end{align}
	Note that $b_n^{(0)}$ are singular in the extremal limit $y\to1/4$.
	This behavior comes from the fact that the singularity $z=z_-$ merges with $z=0$ in this limit.
	Interestingly, we can show that the coefficient $a_n^{(0)}$ can be written in terms of the hypergeometric function as
	\begin{equation}
		a_n^{(0)}=\frac{\left(\frac{1}{6}\right)_n\left(\frac{5}{6}\right)_n}{(n!)^2}\left(\frac{27}{4}\right)^n{}_2F_1\left[\hgarg{-n,-n+1/2}{-3n+1/2};4y\right], \label{an0}
	\end{equation}
	where the Pochhammer symbol is defined as $(x)_n:=\Gamma(x+n)/\Gamma(x)$.
	The proof is presented in Appendix \ref{2F1rec}.

	Around $z=0$, $\omega_1$ and $\omega_2$ can be written as linear combinations of $\omega_r^{(0)}$ and $\omega_l^{(0)}$.
	Since $u_1$ and $u_2$ merge into $0$ in the weak deflection limit ($z\to0$), $\omega_2$ diverges while $\omega_1$ remains finite there.
	Therefore we can set
	\begin{equation}
		\omega_1=A^{(0)}_r\omega^{(0)}_r,\ \ 
		\omega_2=B^{(0)}_r\omega^{(0)}_r+B^{(0)}_l\omega_l^{(0)},
	\end{equation}
	where $A_{r}^{(0)}$ and $B_{r,l}^{(0)}$ are independent of $z$ while can be arbitrary functions of $y$ according to eq.(\ref{zeqomega}).
	However, the $y$-dependences of these coefficients are fixed since $\omega$ must obey the other equation (\ref{yeqomega}) at the same time.
	For $\omega_1$, by taking the leading order term in eq.(\ref{yeqomega}) we can easily see that $\partial_yA_r^{(0)}=0$.
	Similarly we can derive differential equations for $B_r^{(0)}$ and $B_l^{(0)}$ but we don't show them explicitly because
	what we need to describe the deflection angle $\Delta\phi$ is only the finite solution at $z=0$, namely $\omega_1$.

	Among the regular singularities of eq.(\ref{zeqomega}), $z=z_+$ corresponds to the strong deflection limit.
	Thus, we next consider the local solutions around $z=z_+$.
	Since the characteristic exponent is $0$ again, there exists the following basis of the solutions:
	\begin{align}
		\omega_r^{(+)}&=\sum_{n=0}^\infty a_n^{(+)}(z_+-z)^n, \label{omegar+}\\
		\omega_l^{(+)}&=-\omega_r^{(+)}\log(z_+-z)+\sum_{n=1}^\infty b_n^{(+)}(z_+-z)^n,\label{omegal+}
	\end{align}
	where we take a branch of $\log$ so that $\omega_l^{(+)}$ is real on the interval $0<z<z_+$ and the coefficients are normalized by $a_0^{(+)}=1$.
	By substituting these expressions into eq.(\ref{zeqomega}), the coefficients $a_n^{(+)}$ and $b_n^{(+)}$ for $n\geq1$ are uniquely determined as
	\begin{align}
		a_1^{(+)}&=\frac{3(s+1)^3}{512s^3}(36s^2-21s+5),\ \ 
		a_2^{(+)}=\frac{315(s+1)^6}{1048576s^6}\left(432s^4-432s^3+243s^2-78s+11\right), \cdots\\
		b_1^{(+)}&=-\frac{3(s+1)^3}{256s^3(3s+1)}\left(54s^4+216s^3-45s^2-30s+13\right), \\
		b_2^{(+)}&=\frac{27(s+1)^6}{1048576s^6(3s+1)^2}\left(7776s^8-57024s^7-204336s^6 \right. \notag\\
		&\hspace{4em}\left.
			+69120s^5+20673s^4-28116s^3+7014s^2+1548s-719\right),\cdots
	\end{align}
	Recall the definition of $s$, i.e. $s=\sqrt{1-32y/9}$.
	We found the following analytic expression for $a_n^{(+)}$:
	\begin{equation}
		a_n^{(+)}=\frac{\left(\frac{1}{4}\right)_n\left(\frac{3}{4}\right)_n}{(n!)^2}\left(\frac{9(s-1)(s+1)^3}{8s^2}\right)^n
			{}_2F_1\left[\hgarg{-n,2n+1/2}{n+1};\frac{3s+1}{9s(1-s)}\right], \label{anplus}
	\end{equation}
	which can be proved similarly with the proof of (\ref{an0}).

	The strong deflection limit $z\to z_+$ corresponds to the confluent limit of $u_2$ and $u_3$, 
	therefore $\omega_1$ is divergent while $\omega_2$ remains finite in contrast to the weak deflection limit.
	To see the asymptotic form of $\omega_1$ in this limit, we define the connection coefficients $C_r$ and $C_l$ between $z=0$ and $z=z_+$ as
	\begin{equation}
		\omega_r^{(0)}=C_r\omega_r^{(+)}+C_l\omega_l^{(+)}.\label{connection0+}
	\end{equation}
	By inserting this expression into eq.(\ref{yeqomega}), we can obtain the differential equations for $C_r$ and $C_l$ with respect to $y$, or equivalently $s$.
	Isolating the coefficient of $\log(z_+-z)$, we obtain
	\begin{equation}
		\partial_sC_l+\frac{C_l}{2s(s+1)}=0 \ \ \Rightarrow \ \ C_l(s)=\tilde{C}_l\sqrt{\frac{s+1}{s}},\label{Cl}
	\end{equation}
	where $\tilde{C}_l$ is an $s$-independent constant.
	Similarly, taking $O(1)$ term we find 
	\begin{equation}
		\partial_s\left(\sqrt{\frac{s}{s+1}}C_r\right)+\frac{3(s^2-2s-1)}{s(s+1)(3s+1)}\tilde{C}_l=0,
	\end{equation}
	from which we obtain
	\begin{equation}
		C_r=\sqrt{\frac{s+1}{s}}\left[\tilde{C}_l\log\left\{\left(\frac{s}{s+1}\right)^3\frac{1}{3s+1}\right\}+\tilde{C}_r\right], \ \ (\tilde{C}_r \ \text{is $s$-independent}).\label{Cr}
	\end{equation}
	The remaining $s$-independent constants $\tilde{C}_{r,l}$ can be fixed by considering, for example, the neutral limit $s\to1$.
	Eq.(\ref{zeqomega}) reduces to the Gauss hypergeometric equation, namely eq.(\ref{Schwarzschildeq}) with the inhomogeneous term removed.
	Its holomorphic solution at $z=0$ is given by the hypergeometric series\cite{Sasaki_2021},
	\begin{equation}
		\left.\omega_r^{(0)}\right|_{s=1}={}_2F_1\left[\hgarg{1/6,5/6}{1};\frac{27z}{4}\right].
	\end{equation}
	Note that in this case, the strong deflection limit corresponds to $z\to4/27$.
	The analytic continuation of ${}_2F_1$ is well known \cite{HTF1}, from which we obtain
	\begin{equation}
		\left.\omega_r^{(0)}\right|_{s=1}=-\frac{1}{2\pi}\log\left(\frac{1-27z/4}{432}\right)+o(1), \ \ z\to\frac{4}{27}-0.
	\end{equation}
	On the other hand, from eqs.(\ref{omegar+}), (\ref{omegal+}), (\ref{connection0+}), (\ref{Cl}), and (\ref{Cr}), we find 
	\begin{equation}
		\left.\omega_r^{(0)}\right|_{s=1}=\sqrt{2}\left[\tilde{C}_r-\tilde{C}_l\log\left\{\frac{128}{27}\left(1-\frac{27z}{4}\right)\right\}\right]+o(1),\ \ z\to\frac{4}{27}-0.
	\end{equation}
	By equating the above two expressions, we obtain
	\begin{equation}
		\tilde{C}_l=\frac{1}{2\sqrt{2}\pi},\ \ \tilde{C}_r=\frac{\log2048}{2\sqrt{2}\pi}.
	\end{equation}
	Finally, the analytic continuation of $\omega_r^{(0)}$ around $z=z_+$ is given by
	\begin{align}
		\omega_r^{(0)}&=\frac{1}{\pi}\sqrt{\frac{s+1}{8s}}\left[\omega_r^{(+)}\log\left(\frac{2048s^3}{(s+1)^3(3s+1)}\right)+\omega_l^{(+)}\right] \\
		&=-\frac{1}{\pi}\sqrt{\frac{s+1}{8s}}\log\left[\frac{(3s+1)^2}{6912s^3}\left(1-\frac{z}{z_+}\right)\right]+o(1),\ \ z\to z_+-0. \label{omegar0strongleading}
	\end{align}

	The remaining task is to determine the normalization of $\omega_1$, namely $A_r^{(0)}$.
	Since $A_r^{(0)}$ is $y$-independent, we evaluate the integral in eq.(\ref{periods}) at $y=0$ in the limit $z\to+0$.
	By changing the variable of integration to $v=u/\sqrt{z}$, we find
	\begin{equation}
		A_r^{(0)}=\lim_{z\to+0}\omega_1=\lim_{z\to+0}\left.\omega_1\right|_{y=0}=\int_{-1}^1\frac{dv}{\sqrt{1-v^2}}=\pi.
	\end{equation}

	%%}}}
	\subsection{Inhomogeneous Solution\label{inhomsolsection}}
	%%{{{
	Now we consider the inhomogeneous equations (\ref{zeqphi}) and (\ref{yeqphi}).
	In particular, we investigate the local behavior of the solution around $z=0$ and $z=z_+$, which correspond to the weak and strong deflection respectively.
	Since the inhomogeneous term in eq.(\ref{zeqphi}) is $O(z^{-1/2})$ around $z=0$, we can specify the inhomogeneous solution $\Delta\phi_I$ that can be written as
	\begin{equation}
		\Delta\phi_I=\sum_{n=0}^\infty I_n^{(0)}z^{n+1/2}.
	\end{equation}
	From eq.(\ref{zeqphi}), all the coefficients $I_n^{(0)}$ are determined recursively and the first few ones are given by
	\begin{equation}
		I_0^{(0)}=1,\ \ I_1^{(1)}=-\frac{4}{3}(3y-2),\ \ I_2^{(2)}=\frac{8}{5}\left(10y^2-20y+7\right),\cdots.
	\end{equation}
	In fact, one can see the analytic expression for the coefficients $I_n$ is given by
	\begin{equation}
		I_n^{(0)}=\frac{\left(\frac{2}{3}\right)_n(1)_n\left(\frac{4}{3}\right)_n}{n!\left(\frac{3}{2}\right)_n\left(\frac{3}{2}\right)_n}\left(\frac{27}{4}\right)^n{}_2F_1\left[\hgarg{-n,-n-1/2}{-3n-1};4y\right]. \label{Inhomcoeff}
	\end{equation}

	On the other hand, the local behavior of $\Delta\phi_I$ at $z=z_+$ has an ambiguity of adding the homogeneous solutions $\omega_r^{(+)}$ and $\omega_l^{(+)}$
	due to the analytic continuation from $z=0$.
	Since the inhomogeneous term is holomorphic at $z=z_+$, we can expand $\Delta\phi_I$ at $z=z_+$ as
	\begin{equation}
		\Delta\phi_I=-\mathcal{L}_0\sqrt{\frac{s+1}{8s}}\omega^{(+)}_r\log(z_+-z)+\sum_{n=0}^\infty I_n^{(+)}(z_+-z)^n,\label{inhomstrong}
	\end{equation}
	where $\mathcal{L}_0$ is an $s$-independent numerical factor while $I_0^{(+)}, I_1^{(+)},\cdots$ are $s$-dependent in general.
	Note that $\sqrt{(s+1)/8s}$ is factored out so that the coefficient of $\log(z_+-z)$ is a homogeneous solution, see eq.(\ref{Cl}).
	Inserting this expression into eq.(\ref{yeqphi}) and picking the leading order term in the limit $z\to z_+$, we obtain
	\begin{equation}
		\partial_s\left(\sqrt{\frac{s}{s+1}}I_0^{(+)}\right)=-\frac{3\mathcal{L}_0}{2\sqrt{2}}\frac{s^2-2s-1}{s(s+1)(3s+1)}-\sqrt{\frac{3}{8s(3s+1)}}.
	\end{equation}
	Integrating this equation, the $s$-dependence of $I_0^{(+)}$ is determined,
	\begin{equation}
		I_0^{(+)}(s)=\sqrt{\frac{s+1}{8s}}\left[\log\left(\sqrt{3s+1}-\sqrt{3s}\right)^2-\mathcal{L}_0\log\left\{\left(\frac{s+1}{s}\right)^3(3s+1)\right\}
			+\mathcal{I}_0\right], \label{dphiIstrongcoeff}
	\end{equation}
	where $\mathcal{I}_0$ is an $s$-independent constant.
	The remaining constants $\mathcal{L}_0$ and $\mathcal{I}_0$ can be fixed by taking, for example, the neutral limit $s\to 1$. 
	As shown in \cite{Sasaki_2021}, the inhomogeneous solution for the case of $s=1\ (\Leftrightarrow y=0)$ is written in terms of a generalized hypergeometric function,
	\begin{equation}
		\left.\Delta\phi_I\right|_{s=1}=\sqrt{z}{}_3F_2\left[\hgarg{2/3,1,4/3}{3/2,3/2};\frac{27z}{4}\right].
	\end{equation}
	Note that this expression can also be derived from eq.(\ref{Inhomcoeff}) taking the limit $y\to0$.
	The analytic continuation of this generalized hypergeometric function around $z=z_+=4/27$ is known\cite{Buhring1992} and we obtain 
	\begin{equation}
		\left.\Delta\phi_I\right|_{s=1}=-\frac{1}{4}\log\left(\frac{1-27z/4}{432(2-\sqrt{3})^4}\right)+o(1),\ \ z\to z_+-0.
	\end{equation}
	Comparing it with (\ref{inhomstrong}) and (\ref{dphiIstrongcoeff}) at $s=1$, we obtain
	\begin{equation}
		\mathcal{L}_0=\frac{1}{2},\ \ \mathcal{I}_0=\frac{1}{2}\log 2048.
	\end{equation}
	Higher order coefficients $I_n^{(+)}$ for $n\geq1$ are uniquely determined by (\ref{zeqphi}) with the initial condition given above.
	The result can be written as
	\begin{equation}
		I_n^{(+)}=\mathcal{L}_0\sqrt{\frac{s+1}{8s}}b_n^{(+)}+I_0^{(+)}a_n^{(+)}+\tilde{I}_n^{(+)},\ \ n=1,2,3,\cdots,
	\end{equation}
	where the first few terms of $\tilde{I}_n^{(+)}$ are given by
	\begin{align}
		\tilde{I}_1^{(+)}&=\frac{3^{3/2}(s+1)^{7/2}(18s^3+9s^2-16s+5)}{2^{19/2}s^3\sqrt{3s+1}}, \\
		\tilde{I}_2^{(+)}&=-\frac{3^{5/2}(s+1)^{13/2}}{2^{41/2}s^6(3s+1)^{3/2}}\left(3888s^7-29160s^6-13230s^5+25047s^4
			-13392s^3+1722s^2+1190s-385\right).
	\end{align}
	Though we have not succeeded in obtaining an analytic expression for $\tilde{I}_n^{(+)}$, one can calculate up to an arbitrary finite order
	by using the following recurrence relation:
	\begin{equation}
		\sum_{i=0}^{i_{\rm max}}L_i(n-i)\tilde{I}_n^{(+)}=K_n,\ n=1,2,3,\cdots,
	\end{equation}
	where
	\begin{align}
		i_{\rm max}&={\rm min}(3,n), \\
		L_0(n)&=\frac{32s^5(3s+1)^2}{(s+1)^5}n^2,\\
		L_1(n)&=\frac{3s^2(3s+1)}{16(s+1)^2}\bigl[36(9s^4-30s^3+8s^2-2s-1)n^2-36(3s^4+18s^3-4s^2-2s+1)n \notag \\
		&\hspace{4em}
		-(3s+1)(36s^2-21s+5)\bigr], \\
		L_2(n)&=\frac{81(s^2-1)}{1024}\bigl[36(9s^6-96s^5+61s^4-32s^3-5s^2-1)n^2-576s^2(s-1)^2(3s+1)n \notag\\
		&\hspace{4em}
		-(3s+1)(108s^4-243s^3+117s^2+3s-1)\bigr], \\
		L_3(n)&=\frac{19683}{32768}(s^2-1)^4(3s^2+1)(4n+1)(4n+3),
	\end{align}
	and $K_n$ is defined by the $(n-1)$-th order coefficients of the series expansion of the inhomogeneous term in eq.(\ref{zeqphi}) around $z=z_+$,
	\begin{equation}
		-\frac{R_0}{z^{3/2}}=\sum_{n=1}^\infty K_{n}(z_+-z)^{n-1}.
	\end{equation}
	Finally, combining the results so far, we obtain the series expansion of $\Delta\phi_I$ around $z=z_+$,
	\begin{equation}
		\Delta\phi_I=-\frac{1}{2}\sqrt{\frac{s+1}{8s}}\omega_r^{(+)}\log\left\{\left(\frac{s+1}{s}\right)^3\frac{3s+1}{2048}
			\frac{z_+-z}{(\sqrt{3s+1}-\sqrt{3s})^4}\right\}+\sum_{n=1}^\infty\left(\frac{1}{2}\sqrt{\frac{s+1}{8s}}b_n^{(+)}
			+\tilde{I}_n^{(+)}\right)(z_+-z)^n.
	\end{equation}

	%%}}}
	\subsection{Deflection Angle as a Solution to the Inhomogeneous Picard-Fuchs Equations\label{phisolsection}}
	%%{{{

	The deflection angle $\Delta\phi$ is a specific solution to the inhomogeneous PF equations (\ref{zeqphi}) and (\ref{yeqphi}),
	which in general can be written as a sum of a linear combination of the homogeneous solutions $\omega_1$ and $\omega_2$, and the inhomogeneous solution $\Delta\phi_I$.
	From the boundary condition $\Delta\phi(z=0)=\pi/2$ and the discussion in subsections \ref{homsolsection} and \ref{inhomsolsection}, we can conclude 
	\begin{equation}
		\Delta\phi=\frac{1}{2}\omega_1+\Delta\phi_I.
	\end{equation}
	
	The local behaviors of $\Delta\phi$ can be derived from the various expansions of $\omega_1$ and $\Delta\phi_I$ obtained in the previous subsections.
	In the weak deflection limit $(z\to0)$, both terms have power series expansions and the analytic expression is given by 
	\begin{align}
		\Delta\phi&=\frac{\pi}{2}\sum_{n=0}^\infty\frac{\left(\frac{1}{6}\right)_n\left(\frac{5}{6}\right)_n}{n!(1)_n}
			\left(\frac{27z}{4}\right)^n{}_2F_1\left[\hgarg{-n,-n+1/2}{-3n+1/2};4y\right] \notag \\
		&\hspace{4em}
			+z^{1/2}\sum_{n=0}^\infty\frac{\left(\frac{2}{3}\right)_n(1)_n\left(\frac{4}{3}\right)_n}{n!\left(\frac{3}{2}\right)_n
			\left(\frac{3}{2}\right)_n}\left(\frac{27z}{4}\right)^n{}_2F_1\left[\hgarg{-n,-n-1/2}{-3n-1};4y\right],\label{weak_series}
	\end{align}
	or in terms of the original variables,
	\begin{align}
		\Delta\phi&=\frac{\pi}{2}\sum_{n=0}^\infty\frac{\left(\frac{1}{6}\right)_n\left(\frac{5}{6}\right)_n}{n!(1)_n}
			\left(\frac{27M^2}{b^2}\right)^n{}_2F_1\left[\hgarg{-n,-n+1/2}{-3n+1/2};\frac{Q^2}{M^2}\right] \notag \\
		&\hspace{4em}
			+\frac{2M}{b}\sum_{n=0}^\infty\frac{\left(\frac{2}{3}\right)_n(1)_n\left(\frac{4}{3}\right)_n}{n!\left(\frac{3}{2}\right)_n
			\left(\frac{3}{2}\right)_n}\left(\frac{27M^2}{b^2}\right)^n{}_2F_1\left[\hgarg{-n,-n-1/2}{-3n-1};\frac{Q^2}{M^2}\right].\label{weak_series_original}
	\end{align}

	In order to compare this with the result in \cite{Keeton_2005}, we write down the explicit expression for $\alpha=2\Delta\phi-\pi$ up to $O((M/b)^3)$ term,
	\begin{equation}
		\alpha=\frac{4M}{b}+\frac{3\pi}{4}\left(5-\frac{Q^2}{M^2}\right)\left(\frac{M}{b}\right)^2+\left(\frac{128}{3}-\frac{16Q^2}{M^2}\right)\left(\frac{M}{b}\right)^3+O((M/b)^4).
	\end{equation}
	This is identical to the results shown in \cite{Keeton_2005} and the leading behavior is exactly the same as Einstein's formula eq.(\ref{EinsteinFormula}).

	The effect of the background electric charge $y$ appears in higher-order terms. 
	$y$-dependence of $\Delta\phi$ is described by the polynomials ${}_2F_1[-n,-n+1/2;-3n+1/2;4y]$ and ${}_2F_1[-n,-n-1/2;-3n-1;4y]$ for each order of $z$.
	One can see that in the physical region of the parameter $0\leq y\leq 1/4$, these hypergeometric functions are positive and monotonically decreasing.
	This explicitly shows that the deflection angle is a monotonically decreasing function of the background electric charge for a fixed mass and an impact parameter, 
	i.e. $\Delta\phi(z,y_1)\geq\Delta\phi(z,y_2)$ for $0\leq\forall y_1\leq \forall y_2\leq1/4$.
	
	We examine the convergence of (\ref{weak_series}).
	The radius of convergence $R$ for $z$ is given by $R=(\lim a_{n+1}/a_n)^{-1}$, where 
	\begin{equation}
		a_n=\begin{cases}
			\dfrac{\left(\frac{1}{6}\right)_n\left(\frac{5}{6}\right)_n}{(n!)^2}\left(\dfrac{27}{4}\right)^n
			{}_2F_1\left[\hgarg{-n,-n+1/2}{-3n+1/2};4y\right] \ \ (\text{for the homogeneous solution}) \\
			\\
			\dfrac{\left(\frac{2}{3}\right)_n(1)_n\left(\frac{4}{3}\right)_n}{n!\left(\frac{3}{2}\right)_n\left(\frac{3}{2}\right)_n}
			\left(\dfrac{27}{4}\right)^n{}_2F_1\left[\hgarg{-n,-n-1/2}{-3n-1};4y\right] \ \ (\text{for the inhomogeneous solution})
		\end{cases} \label{weak_coeff}
	\end{equation}
	In order to examine the limit $n\to\infty$, we use the following asymptotic expansion of ${}_2F_1$ studied in \cite{Cvitkovicetal2017}, 
	\begin{equation}
		{}_2F_1\left[\hgarg{a-\varepsilon\lambda,b-\lambda}{c};Z\right]\sim\frac{\Gamma(c)(\varepsilon\lambda)^{1/2-c}}
		{\sqrt{2\pi|\varepsilon-1|\sigma}}\frac{(1-t_-)^{c-a+\varepsilon\lambda}}{(-t_-)^{\varepsilon\lambda-a}(1-Zt_-)^{b-\lambda}}\ \ (\lambda\to\infty),
	\end{equation}
	where
	\begin{equation}
		t_\pm=\frac{1-\varepsilon}{2}\pm\frac{|1-\varepsilon|}{2}\sigma,\ \ 
		\sigma=\sqrt{1+\frac{4\varepsilon}{(\varepsilon-1)^2Z}}.
	\end{equation}
	Before applying this formula to eq.(\ref{weak_coeff}), we use the analytic continuation formula 2.10 (3) in \cite{HTF1},
	\begin{align}
		{}_2F_1\left[\hgarg{a,b}{c};x\right]&=B_1(1-x)^{-a}{}_2F_1\left[\hgarg{a,c-b}{a-b+1};(1-x)^{-1}\right] \notag \\
		&+B_2(1-x)^{-b}{}_2F_1\left[\hgarg{b,c-a}{b-a+1};(1-x)^{-1}\right],
	\end{align}
	where 
	\begin{equation}
		B_1=\frac{\Gamma(c)\Gamma(b-a)}{\Gamma(b)\Gamma(c-a)},\ \ 
		B_2=\frac{\Gamma(c)\Gamma(a-b)}{\Gamma(a)\Gamma(c-b)}.
	\end{equation}
	Then, we find the asymptotic behavior of ${}_2F_1$ as
	\begin{align}
		&{}_2F_1\left[\hgarg{-n,-n+1/2}{-3n+1/2};4y\right]\sim\sqrt{\frac{1+s}{2s}}\left(\frac{(1+s)^3}{2(3s+1)}\right)^n, \\
		&{}_2F_1\left[\hgarg{-n,-n-1/2}{-3n-1};4y\right]\sim\frac{(s+1)^2}{2\sqrt{s(3s+1)}}\left(\frac{(1+s)^3}{2(3s+1)}\right)^n \ \ (n\to\infty).
	\end{align}
	Finally, the radii of convergence $R$ are found to be the same and given by
	\begin{equation}
		R=\frac{8(3s+1)}{27(s+1)^3}=z_+.
	\end{equation}
	Therefore, the series expansion (\ref{weak_series}) of $\Delta\phi$ is absolutely convergent for $|z|<z_+$ and $0\leq y\leq 1/4$.

	On the other hand, in the strong deflection limit $(z\to z_+)$, we have the following series expansion:
	\begin{equation}
			\Delta\phi=-\sqrt{\frac{s+1}{8s}}\omega_r^{(+)}\log\left\{\frac{(3s+1)^2}{6912s^3}\frac{1-z/z_+}{(\sqrt{3s+1}-\sqrt{3s})^2}\right\} 
			+\sum_{n=1}^\infty\left(\tilde{I}_n^{(+)}+\sqrt{\frac{s+1}{8s}}b_n^{(+)}\right)\left(z_+-z\right)^n.\label{strong_series}
	\end{equation}
	Note that the coefficients of the power series $\omega_r^{(+)}$ defined by eq.(\ref{omegar+}) are analytically given by eq.(\ref{anplus})
	while we have not succeeded in deriving generic expressions for $\tilde{I}_n^{(+)}$ and $b_n^{(+)}$. 
	The leading behavior in the limit $z\to z_+$ is given by
	\begin{equation}
		\Delta\phi=-\sqrt{\frac{s+1}{8s}}\log\left\{\frac{(3s+1)^2}{6912s^3}\frac{1-z/z_+}{(\sqrt{3s+1}-\sqrt{3s})^2}\right\}+O((z_+-z)\log(z_+-z)). \label{leading_strong}
	\end{equation}
	We can confirm this expression is identical to the result of \cite{TsukamotoGong_2017} taking changes of the variables
	$\Delta\phi\to\alpha=2\Delta\phi-\pi$, $z\to b=2M/\sqrt{z}$, and $s\to Q^2/M^2=9(1-s^2)/8$ into account.

	Before closing this subsection, we take the Schwarzschild $(y\to0 \Leftrightarrow s\to1)$ and the extremal $(y\to1/4 \Leftrightarrow s\to1/3)$ limits of eq.(\ref{weak_series}) as a consistency check.
	Recalling one of the defining properties of the Gauss hypergeometric functions that it is normalized to unity at the origin, 
	it is easy to see that in the Schwarzschild limit, $\Delta\phi$ reduces to 
	\begin{equation}
		\left.\Delta\phi\right|_{y=0}=\frac{\pi}{2}{}_2F_1\left[\hgarg{1/6,5/6}{1};\frac{27z}{4}\right]
		+z^{1/2}{}_3F_2\left[\hgarg{2/3,1,4/3}{3/2,3/2};\frac{27z}{4}\right]. \label{Schwarzschildphi}
	\end{equation}
	To derive the expression in the extremal limit $(y\to1/4)$, we use the following identity:
	\begin{equation}
		{}_2F_1\left[\hgarg{-n,b}{c};1\right]=\frac{(c-b)_n}{(c)_n}\ \ (n=0,1,2,3,\cdots),
	\end{equation}
	from which we obtain
	\begin{align}
		{}_2F_1\left[\hgarg{-n,-n+1/2}{-3n+1/2};1\right]&=\frac{(-2n)_n}{(-3n+1/2)_n}=\frac{\left(\frac{3}{4}\right)_n\left(\frac{1}{4}\right)_n}
			{\left(\frac{5}{6}\right)_n\left(\frac{1}{6}\right)_n}\left(\frac{16}{27}\right)^n, \\
		{}_2F_1\left[\hgarg{-n,-n-1/2}{-3n-1};1\right]&=\frac{(-2n-1/2)_n}{(-3n-1)_n}=\frac{\left(\frac{5}{4}\right)_n\left(\frac{3}{4}\right)_n}
			{\left(\frac{4}{3}\right)_n\left(\frac{2}{3}\right)_n}\left(\frac{16}{27}\right)^n.
	\end{align}
	Then, we find
	\begin{equation}
		\left.\Delta\phi\right|_{y=1/4}=\frac{\pi}{2}{}_2F_1\left[\hgarg{3/4,1/4}{1};4z\right]
		+z^{1/2}{}_3F_2\left[\hgarg{5/4,1,3/4}{3/2,3/2};4z\right].
	\end{equation}
	These results were derived in \cite{Sasaki_2021}. 
	In these cases, series expansions both in the weak and the strong deflection limits were explicitly written. 

	%%}}}
	\subsection{Relationship to the Painlev\'e VI equation\label{PVI-PF}}
	%%{{{
	In the previous subsections, we derived the system of differential equations (\ref{zeqphi}) and (\ref{yeqphi}) for $\Delta\phi$
	and solved to obtain series expansions eq.(\ref{weak_series}) and eq.(\ref{strong_series}).
	Even though the system does have a nontrivial solution given by the integral form (\ref{intexpphi}),
	we discuss the integrability condition in this subsection.
	To do so, we consider the homogeneous equations given by eq.(\ref{zeqomega}) and eq.(\ref{yeqomega}).
	If the coefficients $p_1, p_0, q_1$, and $q_0$ are given arbitrarily, this system doesn't allow any solution other than the trivial one $\omega=0$.
	This fact can be seen by considering a local power series solution at some point, say $z=0$.
	Although eq.(\ref{zeqomega}) and eq.(\ref{yeqomega}) individually give recurrence relations for the expansion coefficients,
	these solutions are not consistent with each other unless $p_i$ and $q_i$ obey certain restrictions.

	When $p_i$ and $q_i$ are rational functions of $z$, such an integrable system is called an isomonodromic (or monodromy-preserving) deformation of (\ref{zeqomega}). 
	In particular, the isomonodromic deformation of a 2nd-order Fuchsian ordinary differential equation 
	with 4 regular singular points and one apparent one is described by the Painlev\'e VI equation (PVI),
	see Appendix\ref{PVIreview}. 
	More precisely, the position of the apparent singular point $\lambda$ as a function of one of the regular singular points $t$ solves PVI (\ref{PVI}).
	In our case, the deformation parameter is essentially given by the background electric charge $y$, or equivalently $s=\sqrt{1-32y/9}$,
	and the position of the apparent singular point is given by $\Lambda=(1-4y)/2y(8y-3)$, see subsection \ref{derivePFeq}.

	To see the correspondence, we have to change the independent variable $z$ to put eq.(\ref{zeqomega}) into the standard form where
	three of the regular singular points lie at $0, 1$, and $\infty$.
	Although such coordinate transformations are not unique, 
	we take $z\to w=z/z_+$ so that the physically relevant region $0\leq z\leq z_+$ is mapped to $0\leq w\leq1$.
	With this transformation, (\ref{zeqomega}) reduces to 
	\begin{equation}
		\left(\partial^2_w+\tilde{p}_1\partial_w+\tilde{p}_0\right)\omega=0,
	\end{equation}
	where 
	\begin{align}
		\tilde{p}_1&=\frac{1}{w}+\frac{1}{w-1}+\frac{1}{w-t}-\frac{1}{w-\lambda}, \\
		\tilde{p}_0&=\frac{\kappa}{w(w-1)}+\frac{\lambda(\lambda-1)\mu}{w(w-1)(w-\lambda)}-\frac{t(t-1)H}{w(w-1)(w-t)},\label{pt0def}\\
		\kappa&=\frac{3}{16},\\
		t&=\frac{z_-}{z_+}=\frac{(s+1)^3(3s-1)}{(s-1)^3(3s+1)},\label{tRN}\\
		\lambda&=\frac{\Lambda}{z_+}=\frac{(s+1)^2(3s-1)}{(s-1)(3s^2+1)},\label{lambdaRN}\\
		\mu&=\frac{(1-s)(3s+1)(3s^2+1)}{48s^2(s+1)^2}, \\
		H&=\frac{(1-s)^3(3s+1)(36s^2+21s+5)}{576s^3(s+1)^3}.
	\end{align}
	By comparing these equations with (\ref{Ap1}) and (\ref{Ap0}), we see that (\ref{tRN}) and (\ref{lambdaRN}) describe an algebraic solution
	to PVI (\ref{PVI}) with $\kappa_0=\kappa_1=\theta=0$ and $\kappa^2_\infty=1/4$.
	
	This algebraic solution to PVI is known in the literature\cite{Dubrovin_1996,Dubrovin_1998,Hitchin1995,Lisovyy2014}.
	In \cite{Dubrovin_1998}, algebraic solutions to (\ref{PVI}) with $\kappa_0=\kappa_1=\theta=0$ were classified
	by using the isomonodromic deformation of a two dimensional Fuchsian system of differential equations.
	They showed that any algebraic solution in this case can be obtained by repeated applications of symmetry transformations to one of the five solutions
	characterized by the values of $\kappa_\infty$\footnote{Note that they used $\mu$ defined by $(2\mu-1)^2=\kappa^2_\infty$, which is different from our definition in eq.(\ref{pt0def}).}.
	In fact, (\ref{tRN}) and (\ref{lambdaRN}) can be obtained by applying the transformation described in Lemma 1.7 to the solution ($A_3$) associated with 
	tetrahedron in their work.
	The solution ($A_3$) is associated with the parameter $\kappa^2_\infty=9/4$ and given explicitly by\footnote{The sign of the parametric representation of the independent variable in ($A_3$) is incorrect.}
	\begin{align}
		\lambda_{9/4}&=\frac{(s-1)^2(3s+1)(9s^2-5)^2}{(1+s)(25-207s^2+1539s^4+243s^6)}, \\
		t&=\frac{(s-1)^3(3s+1)}{(s+1)^3(3s-1)}.
	\end{align}
	As noted in \cite{Dubrovin_1998}, this solution, in an implicit form,  was found in the context of 2-dimensional topological field theories\cite{Dubrovin_1996}.
	From this solution, (\ref{tRN}) and (\ref{lambdaRN}) are generated by the following transformation: 
	\begin{equation}
		\lambda=\frac{\left(\rho_0\dot{\lambda}^2_{9/4}+\rho_1\dot{\lambda}_{9/4}+\rho_2\right)^2\lambda_{9/4}}
			{\sigma_0\dot{\lambda}^4_{9/4}+\sigma_1\dot{\lambda}^3_{9/4}+\sigma_2\dot{\lambda}^2_{9/4}+\sigma_3\dot{\lambda}_{9/4}+\sigma_4},\label{Lemma1.7}
	\end{equation}
	where 
	\begin{align}
		\rho_0&=t^2(t-1)^2,\\
		\rho_1&=-t(t-1)(\lambda_{9/4}-1)(3\lambda_{9/4}-t), \\
		\rho_2&=\frac{\lambda_{9/4}}{4}(\lambda_{9/4}-1)\left[4\lambda_{9/4}(\lambda_{9/4}-1)+4(\lambda_{9/4}-1)(\lambda_{9/4}-t)+(\lambda_{9/4}-t)(\lambda_{9/4}-t-1)\right],\\
		\sigma_0&=t^4(t-1)^4,\\
		\sigma_1&=-4t^3(t-1)^3\lambda_{9/4}(\lambda_{9/4}-1), \\
		\sigma_2&=\frac{1}{2}t^2(t-1)^2\lambda_{9/4}(\lambda_{9/4}-1)\left[12\lambda_{9/4}(\lambda_{9/4}-1)+(\lambda_{9/4}-t)(1+t-3\lambda_{9/4})\right],\\
		\sigma_3&=t(t-1)\lambda^2_{9/4}(\lambda_{9/4}-1)^2\left[-4\lambda_{9/4}(\lambda_{9/4}-1)+(\lambda_{9/4}-t)^2+(\lambda_{9/4}-t)(3\lambda_{9/4}-t-1)\right],\\
		\sigma_4&=\frac{1}{16}\lambda^2_{9/4}(\lambda_{9/4}-1)^2\left[16\lambda^2_{9/4}(\lambda_{9/4}-1)^2-16\lambda_{9/4}(\lambda_{9/4}-1)(\lambda_{9/4}-t)^2\right.
			\notag\\
		&\hspace{4em}\left.
			-8\lambda_{9/4}(\lambda_{9/4}-1)(\lambda_{9/4}-t)(3\lambda_{9/4}-t-1)+(\lambda_{9/4}-t)^2((t-1)^2+\lambda_{9/4}(2t+2-3\lambda_{9/4}))\right],
	\end{align}
	and $\dot{\lambda}_{9/4}=d\lambda_{9/4}/dt$, followed by the reflection of the parameter $s\to-s$.
	The method in \cite{Dubrovin_1998} was extended to the generic PVI and 
	all algebraic solutions were classified in \cite{Lisovyy2014}.
	{\it Solution IV} in this work with a change of the parameter $s\to (s+1)/(s-1)$ is found to be identical to our solution (\ref{tRN}) and (\ref{lambdaRN}).
	We also note that the same solution was obtained in \cite{Hitchin1995}.

	%%}}}

%%}}}
\section{Summary and Discussion\label{summary}}
%%{{{

In this paper, we derived and analyzed the inhomogeneous Picard-Fuchs equations (\ref{zeqphi}) and (\ref{yeqphi}) satisfied by the bending angle $\Delta\phi$
as a function of $z=4M^2/b^2$ and $y=Q^2/4M^2$.
The first equation (\ref{zeqphi}) is shown to be a Fuchsian differential equation with five regular singularities 
including the weak $(z=0)$ and strong $(z=z_+)$ deflection limit and one apparent singularity $z=\Lambda$.
This equation determines the solution space except for possibly $y$-dependent integration constants.
In the Schwarzschild and extremal cases, this equation reduces to inhomogeneous Gauss hypergeometric differential equations. 
Another equation (\ref{yeqphi}) fixes the $y$-dependence of the integration constants as shown in \ref{homsolsection} and \ref{inhomsolsection}.
We investigated local solutions to this system of equations and obtained series expansions of $\Delta\phi$ 
both in the weak (eq.(\ref{weak_series})) and in the strong (eq.(\ref{strong_series})) deflection limits.
In particular, we obtained explicit analytical expressions of the coefficients in the weak deflection limit.
Although we couldn't find analytical expressions of the coefficients in the strong deflection limit,
a linear recurrence relation was derived enabling us to systematically calculate the coefficients up to an arbitrary order.

When we truncate these series expressions at a finite-order term as an approximation for the exact bending angle, we should evaluate their accuracy.
It has been argued that, for the strong deflection limit, the choice of the expansion variable affects the accuracy of the bending angle and related observables\cite{Iyer_2007,Tsukamoto_2022,Tsukamoto_2023}\footnote{The author thanks Naoki Tsukamoto for letting him know this point.}.
Usually, the strong deflection limit is discussed in terms of the variable $b/b_c-1$ as shown in eq.(\ref{BozzaFormula}).
However, it was shown that the 0th-order term of the affine perturbation series using the variable $b'=1-b_c/b$ is more accurate than eq.(\ref{BozzaFormula}) for the cases of a Schwarzschild black hole\cite{Iyer_2007} and an extremal Reissner-Nordstr\"om black hole\cite{Tsukamoto_2022,Tsukamoto_2023}.
Since our variable $1-z/z_+=1-b_c^2/b^2$ in eq.(\ref{leading_strong}) is different from the variables used in eq.(\ref{BozzaFormula}) and in \cite{Iyer_2007,Tsukamoto_2022,Tsukamoto_2023}, 
it should be clarified which expression is more accurate.

The series expressions for $\Delta\phi$ obtained so far are by themselves useful for considering the gravitational lensing problems
by a spherically symmetric and electrically charged massive object.
In addition to this, our method for investigating the deflection angle should be applicable to cases of other types of background metrics.
For example, the deflection angle on the equatorial plane in a rotating black hole spacetime such as the Kerr, or Kerr-Newman spacetime
has an integral expression similar to eq.(\ref{intexpphi}).
We remark that the fact that the deflection angle as a function of background parameters and the impact parameter obeys linear differential equations
would be significant to study series expansions around the weak and strong deflection limit in other spacetimes
since the linear differential equations lead to linear recurrence relations for the expansion coefficients,
which allow to systematically calculate the coefficients up to an arbitrary finite order.
This is in contrast to the method employed in the literature, where various asymptotic behaviors are derived by expanding the integrand 
in power series with a suitable choice of the expansion parameter.

Furthermore, our method can be extended to the cases where the observer and/or the source of photons are located at finite distances from the lensing object\cite{Beloborodov2002,Perlick2004,Beachley2018,GrallaLupsasca2020a,GrallaLupsasca2020b}.
If the source, for example, lies at $r=r_S$ while the observer is kept at infinity, 
the integral expression of the bending angle is given by eq.(\ref{intexpphi}) with the upper limit of the integration replaced by $u=u_S=2M/r_S$.
Note that we here assume the simplest case, where the trajectory doesn't include the point of the closest approach, 
but it is easy to extend to the cases where this assumption doesn't hold.
In this case, the differential equation eq.(\ref{zeqphi}) for the deflection angle becomes
\begin{equation}
	\left(c_2\partial^2_z+c_1\partial_z+c_0\right)\Delta\phi=R(u_S)K^3(u_S)-R(0)K^3(0). \label{zeqphi_finite}
\end{equation}
Observe that the difference between eq.(\ref{zeqphi_finite}) and (\ref{zeqphi}) only appears in the inhomogeneous term $R(u_S)K^3(u_S)$.

We also pointed out that the system of eqs.(\ref{zeqphi}) and (\ref{yeqphi}) can be regarded as an isomonodromic deformation with respect to the parameter $y$ of the Picard-Fuchs equation for the Schwarzschild case,
and the integrability condition for this system is found to be the Painlev\'e VI equation with a specific parameter choice where there exists an algebraic solution.
Isomonodromy implies that the characteristic exponents of the local solutions to the system are independent of the deformation parameter, 
namely the electric charge of the background geometry in this case.
As a result, $\Delta\phi$ is logarithmically divergent in the strong deflection limit for any value of $0\leq y\leq1/4$.
Note that it is known that the divergence of the bending angle in a generic spherically symmetric and asymptotically flat spacetime with a photon sphere is logarithmic\cite{Bozza_2002,Tsukamoto_2017}.

On the other hand, it is known that the logarithmic asymptotic behavior in the strong deflection limit breaks down in some cases.
For example, break down of the logarithmic form was numerically observed for prograde orbits on the equatorial plane in the extremal Kerr black hole in \cite{Bozza_2003},
after which the divergence was analytically shown to be a first-order pole in terms of the impact parameter\cite{Barlow_2017}.
Similarly, one can see that the series expansion (\ref{strong_series}) in the strong deflection limit becomes invalid when $s=0\ (\Leftrightarrow Q^2=9M^2/8)$.
In this case, the spacetime has a naked singularity and the deflection angle was shown to diverge as $1/(b/b_m-1)^{1/6}$\cite{Tsukamoto2020}, 
where $b$ and $b_m$ are the impact parameter and its critical value respectively.
These cases are expected to correspond to singularities of isomonodromic deformation.
In fact, from (\ref{tRN}) and (\ref{lambdaRN}), we see that $s=0$ corresponds to $t=\lambda=1$, which is a singularity of the Painlev\'e VI equation (\ref{PVI}).
Although the series expression (\ref{strong_series}) is useless in this case, the differential equation (\ref{zeqphi}) remains valid.
The differential operator of (\ref{zeqphi}) becomes $c_2\partial_z^2+c_1\partial_z+c_0\to 16z(27z-8)^2\partial_z^2+32(27z-8)(27z-4)\partial_z+3(729z-248)$ in this case,
from which the characteristic exponents at the strong deflection limit $z=8/27$ are found to be $\pm1/6$.
The singular behavior for prograde orbits in the extremal Kerr spacetime would be understood in a similar manner.

Lastly, we discuss a mathematical aspect related to the Painlev\'e equation.
The Reissner-Nordstr{\"{o}}m spacetime has various extensions such as the Kerr-Newman spacetime, which include additional background parameters.
Our method seems to be applicable to these cases as well and 
the resulting systems of differential equations for the bending angle would contain three or more independent variables.
Similarly to the case treated in this article, these equations must obey some integrability conditions, which should be a kind of generalizations of the Painlev\'{e} VI equation.
For example, a multivariable generalization of the Painlev\'{e} VI equation is known as the Garnier system (see for example \cite{Okamoto1986Iso,Gauss2Painleve}).
As far as we know, the classification of all algebraic solutions for the Garnier system is not known,
while various specific cases are known, for example \cite{Tsuda_2003,Suzuki_2006,Kawamuko_2013,DiarraLoray_2020}.
Applying our method to other types of background spacetimes which are generalizations of the Schwarzschild spacetime
may give specific solutions to such equations.

%%}}}
\appendix
\section{Isomonodromic deformation and Painlev\'e VI equation\label{PVIreview}}
%%{{{

In this appendix, we briefly describe the relationship between the Painlev\'e VI equation (PVI) (\ref{PVI}) and the isomonodromic deformation of a Fuchsian differential equation.
See, for example \cite{Gauss2Painleve} for more detail.
PVI is a nonlinear 2nd-order ordinary differential equation, which appears as an integrability condition for an isomonodromic deformation of a Fuchsian equation
with 4 regular singular points and 1 apparent singular point on the complex plane\cite{Okamoto1980-1}.
Usually, problems of isomonodromic deformations for Fuchsian differential equations are discussed by using the corresponding systems of 1st-order equations,
where the integrability conditions are reduced to the so-called Schlesinger's equations.
There are various advantages to using the 1st-order systems. 
For example, no apparent singularity appears in the 1st-order systems even though the corresponding 2nd (or higher) order equations possess
one or more apparent singularities in the problem of isomonodromic deformations.
In spite of this, we directly treat the 2nd-order equations to avoid the complexity of gauge freedom introduced by considering the 1st-order system;
the 1st-order systems corresponding to a given 2nd-order equation are not unique.
For other aspects related to the Painlev\'e equations such as the Hamiltonian structure or extension to the cases with more singularities, 
see for example \cite{Okamoto1980-1,Okamoto1980-2,Okamoto1986Iso,Clarkson2003,Guzzetti_2015}.

We begin with the following 2nd-order Fuchsian differential equation:
\begin{equation}
	\left[\partial_z^2+p_1\partial_z+p_0\right]\omega=0. \label{2ndODE}
\end{equation}
where $p_1$ and $p_0$ are rational functions of $z$.
We restrict ourselves to the cases where (\ref{2ndODE}) has 5 regular singularities, 
where $p_1$ and $p_0$ have at most 1st and 2nd-order poles respectively.
We can always change the positions of these singularities by means of the linear fractional transformation with respect to $z$ so that 
any three of the regular singular points are located at $z=0, 1$, and $\infty$.
We denote the positions of the remaining singularities by $z=t$
\footnote{$t$ here is different from the spacetime coordinate and represents just a deformation parameter.} and $z=\lambda$ respectively.
We next assume $p_1$ and $p_0$ take the following form:
\begin{align}
	p_1&=\frac{1-\kappa_0}{z}+\frac{1-\kappa_1}{z-1}+\frac{1-\theta}{z-t}-\frac{1}{z-\lambda},\label{Ap1}\\
	p_0&=\frac{\kappa}{z(z-1)}+\frac{\lambda(\lambda-1)\mu}{z(z-1)(z-\lambda)}-\frac{t(t-1)H}{z(z-1)(z-t)},\label{Ap0}
\end{align}
where $\kappa_0,\ \kappa_1,\ \theta,\ \mu$, and $H$ are $z$-independent.
Note that $p_0$ is assumed to have no 2nd-order pole since such a term can always be eliminated by a suitable transformation of the dependent variable $\omega$.
In this form, we can easily read off the local behavior of the solutions at each singular point from the residue of $p_1$.
For example, the characteristic exponents at $z=0$ are $0$ and $\kappa_0$.
Similarly one can see that the characteristic exponents at $z=\lambda$ are $0$ and $2$.
Although the difference of the characteristic exponents is an integer,
we here require any local solution at $z=\lambda$ has no logarithmic term.
Then, the following restriction is derived:
\begin{equation}
	t(t-1)H=(\lambda-t)\kappa+\lambda(\lambda-1)(\lambda-t)\mu^2
		-\mu\left\{(\lambda-1)(\lambda-t)\kappa_0+\lambda(\lambda-t)\kappa_1+\lambda(\lambda-1)(\theta-1)\right\}. \label{Hdef}
\end{equation}
In this case, $z=\lambda$ is called an apparent singularity.

Now we regard $t$, one of the positions of the regular singularities, as a parameter to deform the differential equation (\ref{2ndODE}).
If we require that the monodromy of the original differential equation (\ref{2ndODE}) is independent of $t$, 
it is known that there must exist rational functions $q_1$ and $q_0$ of $z$ and the solution $\omega$ obeys
\begin{equation}
	\left[\partial_t+q_1\partial_z+q_0\right]\omega=0. \label{deformeq}
\end{equation}

Because $\kappa_0,\ \kappa_1$, and $\theta$ determine the characteristic exponents at the regular singularities,
which determines the local monodromy of (\ref{2ndODE}), they must be $t$-independent constants.
On the other hand, $\lambda, \ \mu$ and $H$ can be functions of $t$ under the constraint (\ref{Hdef}).

In order for the system of differential equations (\ref{2ndODE}) and (\ref{deformeq}) to allow nontrivial solutions,
$p_i$ and $q_i$ must satisfy some integrability conditions.
The integrability conditions can be derived by equating $\partial_t\partial^2_z\omega$ and $\partial^2_z\partial_t\omega$ calculated in two ways from (\ref{2ndODE}) and (\ref{deformeq}).
This results in 
\begin{align}
	\partial_tp_1&=\partial_z\left(\partial_zq_1-p_1q_1+2q_0\right), \\
	\partial_tp_0&=\partial^2_zq_0+p_1\partial_zq_0-2p_0\partial_zq_1-q_1\partial_zp_0.
\end{align}
Undetermined functions of $t$ are $\mu(t)$ and $\lambda(t)$ due to the restriction (\ref{Hdef}).
One can show that the above integrability conditions reduce to a system of differential equations for $\mu(t)$ and $\lambda(t)$ and 
in fact these are found to be the following Hamiltonian system:
\begin{equation}
	\frac{d\lambda}{dt}=\frac{\partial H}{\partial\mu}(\lambda,\mu,t),\ \ \frac{d\mu}{dt}=-\frac{\partial H}{\partial\lambda}(\lambda,\mu,t).
\end{equation}
Here, $H$ is regarded as a function of $\lambda, \mu$, and $t$ defined by (\ref{Hdef}).
Eliminating $\mu$ from these equations, $\lambda$ can be shown to satisfy PVI, of which explicit form is given by
\begin{align}
	\frac{d^2\lambda}{dt^2}&=\frac{1}{2}\left(\frac{1}{\lambda}+\frac{1}{\lambda-1}+\frac{1}{\lambda-t}\right)\left(\frac{d\lambda}{dt}\right)^2
		-\left(\frac{1}{t}+\frac{1}{t-1}+\frac{1}{\lambda-t}\right)\frac{d\lambda}{dt} \notag\\
	&\hspace{4em}
		-\frac{\lambda(\lambda-1)(\lambda-t)}{t^2(t-1)^2}\left[\frac{t\kappa^2_0}{2\lambda^2}+\frac{(1-t)\kappa^2_1}{2(\lambda-1)^2}
		+\frac{t(t-1)(\theta^2-1)}{2(\lambda-t)^2}-\frac{1}{2}\kappa^2_\infty\right],\label{PVI}
\end{align}
where $\kappa^2_\infty$ is given by $4\kappa=(\kappa_0+\kappa_1+\theta-1)^2-\kappa^2_\infty$.

%%}}}
\section{Solving the four-term recurrence relation\label{2F1rec}}
%%{{{

By inserting the series expansion (\ref{omega_r0}) into (\ref{zeqomega}), the following four-term recurrence is obtained:
\begin{align}
	&16(n+3)^2(4y-1)^2a_{n+3}^{(0)}\notag\\
	&\hspace{2em}+(4y-1)\left[12(64y^2-56y+9)n^2+4(832y^2-792y+135)n+3632y^2-3720y+663\right]a_{n+2}^{(0)} \notag \\
	&\hspace{2em}
		+2y\left[4(1536y^3-1664y^2+648y-81)n^2+8(2048y^3-1792y^2+648y-81)n\right. \notag \\
	&\hspace{4em}
		\left.+11008y^3-7728y^2+2520y-315\right]a_{n+1}^{(0)}\notag\\
	&\hspace{2em}
		+128y^4(8y-3)(4n+3)(4n+1)a_n^{(0)}=0. \label{an0fourtermrec}
\end{align}
In this appendix, we show that the expression (\ref{an0}) really solves this recurrence by using the contiguous relations for the Gauss hypergeometric functions,
which are linear identities among three hypergeometric functions ${}_2F_1[a,b;c;x]$ whose parameters differ by $\pm1$\cite{HTF1}.
It is known that any contiguous relations can be found by combining 15 fundamental relations \cite{VIDUNAS2003507}.
Among the 15 fundamental contiguous relations, we use the following:
\begin{align}
	0&=a(1-x)F(a+1)+\left(c-2a-(b-a)x\right)F-(c-a)F(a-1) \label{ct1}\\
	0&=(b-a)F+aF(a+1)-bF(b+1) \label{ct2}\\
	0&=(c-a-b)F+a(1-x)F(a+1)-(c-b)F(b-1) \label{ct3}\\
	0&=c\left(a-(c-b)x\right)F-ac(1-x)F(a+1)+(c-a)(c-b)xF(c+1) \label{ct4}\\
	0&=(c-a-1)F+aF(a+1)-(c-1)F(c-1) \label{ct5}\\
	0&=c(1-x)F-cF(a-1)+(c-b)xF(c+1) \label{ct6}\\
	0&=\left(a-1-(c-b-1)x\right)F+(c-a)F(a-1)-(c-1)(1-x)F(c-1).\label{ct7}
\end{align}
Here, $F={}_2F_1[a,b;c;x], F(a+1)={}_2F_1[a+1,b;c;x], F(b+1)={}_2F_1[a,b+1;c;x]$, and so on.
By using these identities, we show that (\ref{an0}) satisfies the four-term recurrence (\ref{an0fourtermrec}).

Since the expression (\ref{an0}) contains ${}_2F_1[-n,-n+1/2;-3n+1/2;4y]$, we have to derive the contiguous relation among $F$ and $F(a\pm1,b\pm1,c\pm3)$.
To do so, we follow the method described in \cite{VIDUNAS2003507};
in order to derive the contiguous relation among $F$, $F(a+k,b+l,c+m)$, and $F(a+p,b+q,c+r)$ for a given set of integers $k,l,m,p,q,r$,
we first express the latter two functions as a linear combinations of $F$ and $F(a+1)$.
Then by eliminating $F(a+1)$ from these expressions, we can obtain the desired identity.
In order to achieve this, we first derive the identity among $F, F(a+1)$, and $F(a+1,b+1)$.
By setting $a\to a+1$ in (\ref{ct1}) and (\ref{ct2}), and eliminating $F(a+2)$ from them, we have
\begin{equation}
	0=(a-c+1)F+(a+b-c+1)F(a+1)+b(1-x)F(a+1,b+1). \label{a+b+}
\end{equation}
From this identity, we next derive the identity among $F, F(a+1)$, and $F(a+1,b+1,c+1)$.
By setting $c\to c+1$ in (\ref{a+b+}) and (\ref{ct6}), and eliminating $F(a+1,c+1)$, we have the identity among $F, F(a+1), F(c+1)$, and $F(a+1,b+1,c+1)$.
Combining this identity with (\ref{ct4}) we can eliminate $F(c+1)$ and obtain
\begin{equation}
	0=cF-cF(a+1)+bF(a+1,b+1,c+1).
\end{equation}
Repeating this process, the identity among $F, F(a+1)$, and $F(a+1,b+1,c+3)$ is found to be
\begin{align}
	0&=c(c+1)(c+2)\left[(c-b)(c-b+1)x^2+\left(b(c+a+1)-2c(c+1)\right)x+c(c+1)\right]F \notag\\
	&\hspace{4em}
	-c(c+1)(c+2)(1-x)\left[c(c+1)(1-x)+abx\right]F(a+1) \notag \\
	&\hspace{6em}
	+b(c-a)(c-b)(c-a+1)(c-b+1)x^3F(a+1,b+1,c+3). \label{a+1b+1c+3}
\end{align}
Similarly, by using (\ref{ct1}), (\ref{ct3}), (\ref{ct5}), and (\ref{ct7}), we obtain the identity among $F, F(a+1)$, and $F(a-1,b-1,c-3)$,
\begin{align}
	0&=(c-a-1)\left[(c-2)(c-3)-(c-a-2)(b-1)x\right]F \notag\\
	&\hspace{4em}
	+a\left[(c-2)(c-3)(1-x)+(a-1)(b-1)x\right]F(a+1) \notag \\
	&\hspace{6em}
	-(c-1)(c-2)(c-3)F(a-1,b-1,c-3).\label{a-1b-1c-3}
\end{align}
Finally, by eliminating $F(a+1)$ from (\ref{a+1b+1c+3}) and (\ref{a-1b-1c-3}), we obtain the identity among $F$ and $F(a\pm1,b\pm1,c\pm3)$,
\begin{align}
	0&=ab(c-a)(c-b)(c-a+1)(c-b+1)x^3\left[(c-2)(c-3)(1-x)+(a-1)(b-1)x\right]F(a+1,b+1,c+3) \notag\\
	&\hspace{2em}
	-(c-3)(c-2)(c-1)c(c+1)(c+2)(1-x)\left[c(c+1)(1-x)+abx\right]F(a-1,b-1,c-3) \notag \\
	&\hspace{2em}
	+c(c+1)(c+2)\left[(c-3)(c-2)(c-1)c(c+1)+(c-2)(c-1)c\left(4ab-a-b+7-(a+b-5)c-2c^2\right)x \right. \notag \\
	&\hspace{2em}
	+(c-1)\left(3(a^2b^2-a^2b-ab^2+ab)+2(8ab-2a-2b+5)c-(8ab+2a+2b-3)c^2+2(a+b-3)c^3+c^4\right)x^2 \notag \\
	&\hspace{2em}
	+\left(ab(a-1)(b-1)(a+b+3)-2(2a^2b^2-2a^2b-2ab^2-ab+a+b-1)c\right.\notag \\
	&\hspace{2em}
		\left.\left.-(12ab-a-b+1)c^2+2(2ab+a+b-1)c^3-(a+b-1)c^4\right)x^3\right]F 
\end{align}

By setting $a=-n, b=-n+1/2$, and $c=-3n+1/2$ and multiplying a suitable factor, we obtain the following three-term recurrence for $a_n^{(0)}$:
\begin{align}
	16(n+1)^2\left[(16n-3)x-3(6n-1)\right](1-x)a_{n+1}^{(0)}-(4n-1)(4n-3)\left[(16n+13)x-3(6n+5)\right]x^3a_{n-1}^{(0)} \notag \\
	-\left[(512n^3+416n^2-9)x^3-3(6n+1)(160n^2+104n-21)x^2\right. \notag \\
	\left.+3(40n+33)(6n-1)(6n+1)x-9(6n+5)(6n-1)(6n+1)\right]a_n^{(0)}=0, \label{3termrec}
\end{align}
where $x=4y$ is understood.
We denote the left side of (\ref{3termrec}) by $R(n)$.
Various four-term recurrences among $a_n^{(0)}, a_{n+1}^{(0)}, a_{n+2}^{(0)}$, and $a_{n+3}^{(0)}$ can be derived by adding $R(n+1)$ and $R(n+2)$ with arbitrary weights.
By direct computation, we found the combination $(1-x)R(n+2)-x(2x-3)R(n+1)/2=0$ is exactly the same as the four-term recurrence (\ref{an0fourtermrec}),
which concludes that the expression (\ref{an0}) solves the recurrence (\ref{an0fourtermrec}).

%%}}}

\printbibliography
\end{document}